\documentclass[twocolumn,amsmath,amssymb,showpacs]{revtex4}
\usepackage{subfigure}
\usepackage{wrapfig}
\usepackage{graphicx}
\usepackage{dcolumn}
\usepackage{bm}
\usepackage{amsmath}
\usepackage{amsfonts}
\usepackage{amssymb}
\usepackage{slashbox}
\makeatletter

\begin{document}
\title{Simulation of granular jet: Is granular flow really a ``perfect fluid?"}
\author{Tomohiko G. Sano}
\author{Hisao Hayakawa}
\affiliation{Yukawa Institute for Theoretical Physics, Kyoto University Kitashirakawa Oiwakecho, Sakyo-ku, Kyoto 606-8502 Japan}
\begin{abstract}
We perform three-dimensional simulations of a granular jet impact for both frictional and frictionless grains. 
Small shear stress observed in the experiment[X. Cheng {\it et al.}, Phys. Rev. Lett. {\bf 99}, 188001 (2007) ] is reproduced through our simulation. 
However, the fluid state after the impact is far from a perfect fluid, and thus, similarity between granular jets and quark gluon plasma is superficial, because the observed viscosity is finite and its value is consistent with the prediction of the kinetic theory. 

\end{abstract}
\maketitle

\section{Introduction} 
Impact processes play important roles in various fields such as nuclear reactions\cite{11,32,34}, atomic collisions\cite{53,55}, hydrodynamics\cite{4,3,25,21} and granular physics\cite{1,45,46,47,40,43,41}. Recent experimental and numerical studies revealed interesting aspects of impact processes of a granular flow. At low volume fractions, impact of a granular flow onto a wall produces a shock, which quantitatively agrees with the Mach cone produced by supersonic gas flow\cite{45,46,47}. The impact dynamics of granular particles is important not only for industrial applications, e.g. ink-jet printing and blast cleaning\cite{56,57}, but for geophysical problems such as formation of craters\cite{40,43,41}. 

Recently, an experimental paper on dense granular jets\cite{1} has reported that the fluid state after the impact is similar to that for quark-gluon plasma (QGP) achieved in heavy ion colliders, 
where QGP behaves as a fluid with very small viscosity\cite{11,32,34}.
Quite recently, Ellowitz {\itshape et al.}\cite{0} demonstrated that
the solution of inviscid Euler equation is almost identical to that obtained from their molecular dynamics simulation for inelastic hard core particles, at least, for two dimensional frictionless grains.
These results are counter intuitive because in a usual setup the dense granular fluid has a large viscosity \cite{54}.

The purpose of this paper is to clarify whether the granular fluid after the impact on a fixed wall actually behaves as a perfect fluid. So far all numerical studies are two-dimensional ones\cite{0,39,8}, we perform a three dimensional molecular dynamics simulation for soft core particles to study fluid states after the impact. The observed shear viscosity is finite and consistent with the result of the kinetic theory for granular flow. However, because the strain rate is small, the observed shear stress is small. Thus, the similarity between QGP,  which is characterized by the small viscosity \cite{32,11}, and granular jet is superficial.

After the introduction of our numerical model in Sec. II, we present the results for scattering of granular jets observed in our simulation to compare them with the experimental results in Ref. \cite{1} in Sec. III. In Sec. IV, we analyze the local stress tensor in the cylindrical coordinate and compare results with the kinetic theory, which reproduces our results for pressure and shear viscosity, except for them near the symmetric axis on the target. We find that large normal stress difference, which is the difference of diagonal components of stress tensor, exists. In Sec V, the origin of anisotropic temperature is discussed and results are summarized in Sec VI.

\begin{figure}[h]
\includegraphics[scale = 0.7]{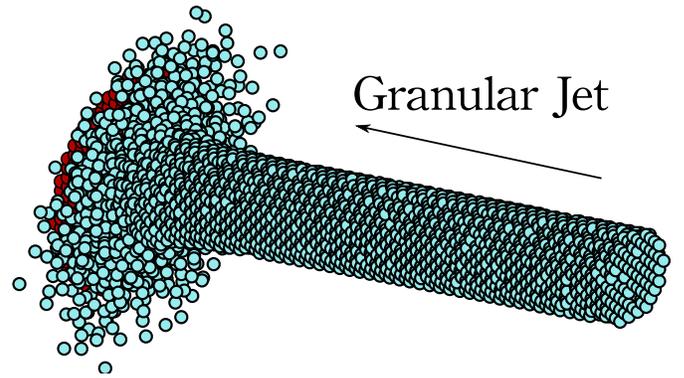}
\caption{(Color online) Snapshot of a three-dimensional simulation. Sky-blue-colored particles are grains and red ones are wall particles. Grains consisting of a regular lattice with missing particles collide on a bumpy wall, where grains are scattered randomly, and the jet ejected along with the wall. }
\label{fig:epsart}
\end{figure}

\section{Model} 
We adopt the discrete element method (DEM) for mono-disperse soft core particles of the diameter $d$\cite{6,7}.
The reason why we adopt a soft core model is the following: 
The DEM can be used even for dense systems above the jamming point, while 
the event driven(ED) algorithm simulation cannot reach the jamming point. We also indicate that  the primitive ED algorithm encounters the inelastic collapse, though it  can be avoided by introducing velocity-dependent restitution coefficient to the ED algorithm\cite{37}. In contrast, DEM has an advantage that it is free from the inelastic collapse and we can easily include the effect of friction and rotation of grains. 
 When the particle $i$ at the position ${\bf r}_i$ and the particle $j$ at ${\bf r}_j$ are in contact, 
the normal force $F^n _{ij}$ is described as $F^n _{ij}  \equiv  F_{ ij} ^{\rm (el)} + F_{ ij} ^{\rm (vis)}$ with  $F_{ ij} ^{\rm (el)}  \equiv  k_n(d-r_{ ij})$ and $F_{ ij} ^{\rm (vis)}  \equiv -\eta_n({\bf g}_{ ij} \cdot \hat{\bf r}_{ ij})$,
 where  $r_{ij}  \equiv  |{\bf r}_{i} - {\bf r}_{j}|$ and  ${\bf g}_{ij}  \equiv  {\bf v}_{ i} - {\bf v}_{ j}$ 
with the velocity ${\bf v}_i$ of the particle $i$.
 The tangential force is given by $F^t _{ij} \equiv \min\{ |\tilde{F^t _{ij}}|, \mu F_{ij}^n \}{\rm sign}(\tilde{F}_{ij}^t) $, where  $\mu$ is the friction constant,
 ${\rm sign}(x)=1$ for $x\ge 0$ and ${\rm sign}(x)=-1$ for otherwise, 
$\tilde{F^t _{ij}} \equiv k_{ t} \delta^t _{ij} - \eta_{ t} \dot{\delta}_{ij} ^t$
with the tangential overlap $\delta^t _{ij}$ between $i$ and $j$ particles and
 the tangential component of relative velocity $\dot{\delta}^t_{ij}$ between $i$ th and $j$ th particles. 
For most cases except for Sec. III A, we adopt parameters $k_t = 0.2 k_n, \eta_t = 0.5 \eta_n$, $\mu = 0.2$, $k_n = 4.98 \times 10^2 mu_0 ^2 /d^2$ and $\eta_n = 2.88 u_0 /d$, with incident velocity $u_0$ and the particle mass $m$. Experimentally, the friction constant of nylon spheres is known to be $\mu = 0.175 \pm 0.1$\cite{59}.
This set of parameters implies that the restitution coefficient for normal impact is ${e} = 0.75$ and duration time is $t_c = 0.10 d/u_0$. 
We adopt the second-order Adams-Bashforth method for the time integration with the time interval $\Delta t = 0.02 t_c$. 

Initial configurations are generated as follows: We prepare fcc crystals and remove particles randomly to reach the desired density. 
We control the initial volume fraction $\phi_0 / \phi_{\rm fcc} \equiv \tilde{\phi_0}$ before the impact as $0.30 \leq \tilde{\phi_0} \leq 0.90$ with volume fraction for a fcc crystal $\phi_{\rm fcc} \simeq 0.74$ and 20,000 particles are used. 
The initial granular temperature, which represents the fluctuation of particle motion, is zero. 
The wall consists of one-layer of particles, which are connected to each other and with their own initial positions via the spring and the dashpot with spring constant $k_p = 10.0 mu_0 ^2 /d^2$ and the dashpot constant  $\eta_p = 5.0 \eta_n$, respectively.  It is known that the collective motion of particles near the wall is almost frozen\cite{0},
while the granular temperature near the wall is higher than that in the other regions for our cases. 

Figure 1 is a snapshot of our simulation on the impact of granular jet. The grains on a regular lattice with missing particles collides on a bumpy wall, where grains are scattered randomly, and the jet ejected along with the wall. 
\section{Results for scattering of granular jets}
\subsection{Scattering angle}
\begin{figure}[h]
\includegraphics[scale = 0.70]{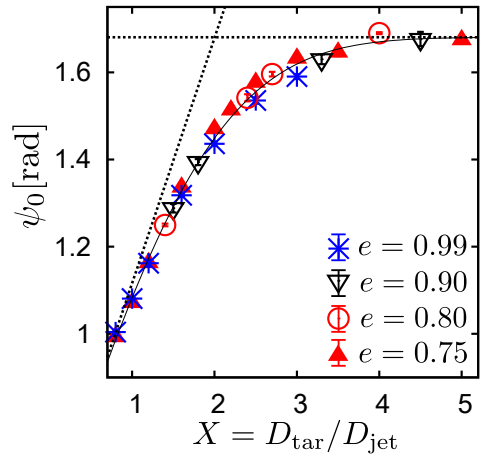}
\caption{(Color online) The dependence of $\psi_0$ [rad] on $X$ for the frictional case with $\tilde{\phi}_0 = 0.90$, $e = 0.75$ and $D_{\rm jet}/d = 4.5$. $\psi_0$ does not depend on the restitution coefficient. The dotted lines are asymptotic ones $\psi_0 = C_0$ for $X \gg 1$ and $\psi_0 = C_1 X + C_2$ for $X \ll 1$ and the solid line is the interpolation function. The asymptotic behavior which is observed in experiments\cite{1} is obtained.}
\label{scatt}
\end{figure}

The scattering angle $\psi_0$ for the frictional case with $\tilde{\phi}_0 = 0.90$, and $D_{\rm jet}/d = 4.5$ for several $e$ exhibits the crossover from a cone-like structure to a sheet-like one, depending on $X \equiv D_{\rm tar}/D_{\rm jet}$, which is almost independent of the restitution coefficient\cite{58}(Fig. \ref{scatt}). We average the data over ten different initial configurations in the followings. The dotted lines in Fig. \ref{scatt}, $\psi_0 = C_0$ for $X \gg 1$ and $\psi_0 = C_1 X + C_2$ for $X \ll 1$, are asymptotic lines, which are expected in Ref. \cite{4} with constants $C_0, C_1$ and $C_2$. The solid line is an interpolation function $\psi_0 = \sqrt{C_0 ^{'2} \{1-\exp(-C_1 ^{2} X^2 / C_0 ^{'2})\}} + C_2$, which reproduces the asymptotic behavior for $\psi_0$ with $C_0 ^{'} \equiv C_0 - C_2$. We obtain fitting parameters $C_0 = 1.68, C_1 = 0.563$ and $C_2 = 0.554$ by fitting the interpolation function to the numerical data.

\subsection{On the effect of the initial spatial anisotropy}
\begin{figure}[h]
\includegraphics[scale = 0.70]{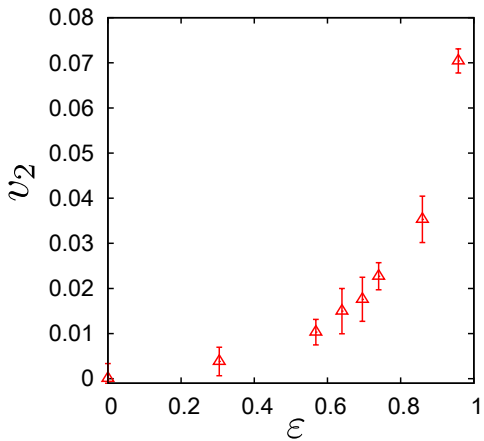}
\caption{(Color online) The dependence of $v_2$ on $\varepsilon$ for the frictional case with $\tilde{\phi}_0 = 0.90$ and $e = 0.75$, where the initial cross section of the jet is fixed to $16d^2$. $v_2$ does not linearly depend on $\varepsilon$. It should be noted that the observed $v_2$ is much smaller than the value reported in Ref. \cite{1}.}
\label{elliptic}
\end{figure}

To estimate the effect of the initial spatial anisotropy, $v_2$, which is the coefficient of $\cos 2 \varphi$ with the azimuthal angle $\varphi$ for scattering flux $dN/d\varphi$, is conventionally used\cite{11}. The scattering flux $dN/d\varphi$ is related to the differential scattering cross section $d\sigma(\theta,\varphi) / d\Omega$ with the scattering angle $\theta$ and the other coefficients $v_n$ with $n = 0,1, \cdots$ as $dN/d\varphi \equiv \int d\sigma / d\Omega d\cos \theta = \sum_{n=0} ^{\infty} v_n \cos n \varphi$. If the fluid behaves as a perfect fluid, $v_2 / \varepsilon$ is expected to be a constant, where $\varepsilon$ is the eccentricity $\varepsilon = (l^2 -1)/(l^2 + 1)$ with an aspect ratio $l$ for an initial cross section of the jet.  The aspect ratio is changed with fixing the area for the cross section to $16d^2$. Although $v_2$ is enhanced as $\varepsilon$ increases, the observed $v_2$ is not proportional to $\varepsilon$ for $e = 0.75$ (Fig. \ref{elliptic}). We also indicate that Ref. \cite{1} only reports one parameter of $v_2 = 0.16$ and $\varepsilon = 0.615$ and thus, they cannot discuss whether $v_2$ is proportional to $\varepsilon$. It should be noted, however, that our $v_2$ is much smaller than the experimental value for the same $\varepsilon$.

\section{Results for Rheology of granular jets}
\begin{figure}[h]
\includegraphics[scale = 0.35]{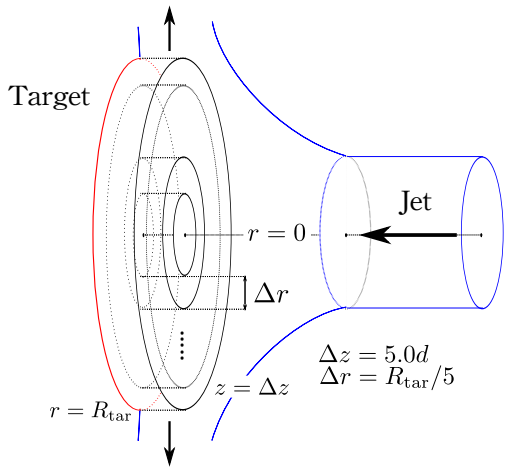}
\caption{(Color online) A schematic picture of the calculation region. The cylinder with radius $R_{\rm tar}$ and height $5d$ is divided into cylindrical mesh, where physical quantities are estimated. }
\label{calc}
\end{figure}

We evaluate physical quantities near the wall at the height $z = \Delta z \equiv 5.0d$ from the wall $z = 0$ for $e = 0.75$. The jet diameter $D_{\rm jet} /d = 10.0$ and the target diameter $D_{\rm tar} / d= 22.0$ are fixed. We divide cylindrical calculation region into the radial direction $r = 0, \Delta r, \cdots, 5\Delta r$, with $\Delta r \equiv R_{\rm tar} /5$ and the target radius $R_{\rm tar}$, and estimate physical quantities in the corresponding mesh region with $k\Delta r < r < (k+1)\Delta r$ ($k = 0,1,\cdots,5$), where $r$ is denoted to the distance from the symmetric axis of the cylindrical coordinate(Fig. \ref{calc}).

\subsection{Stress tensor}
Let us evaluate the stress tensor near the wall as in Ref. \cite{60}. 
The microscopic definition of the stress tensor at ${\bf r}$ is given by
\begin{equation}
\sigma_{\alpha \beta}({\bf r}) = \frac{1}{V} \sum_{i} m u_{i \alpha} u_{i \beta} + \frac{1}{V} \sum_{i<j} F_{\alpha} ^{ij} r_{\beta} ^{ij},
\end{equation}
where $i$ and $j$ are indices of particles, $\alpha, \beta = r,\theta,z$ denotes cylindrical coordinates and $\sum$ denotes the summation over the particles located at ${\bf r}$. 
Here, $z$ axis is parallel to the incident jet axis, and 
$V$ is the volume of each mesh at ${\bf r}$ and $u_{i \alpha} = v_{\alpha} ^{i} - {\bar v}_{\alpha}({\bf r})$ with the mean velocity $ {\bar v}_{\alpha}({\bf r})$ in the mesh at ${\bf r}$.
To calculate the stress tensor in cylindrical coordinates, we firstly calculate $\sigma_{\alpha' \beta'}$ in Cartesian coordinate, $\alpha', \beta' = x,y,z$, whose origin is the same as cylindrical one, and transform it into that for cylindrical one.

\begin{figure}[h]
\includegraphics[scale = 0.65]{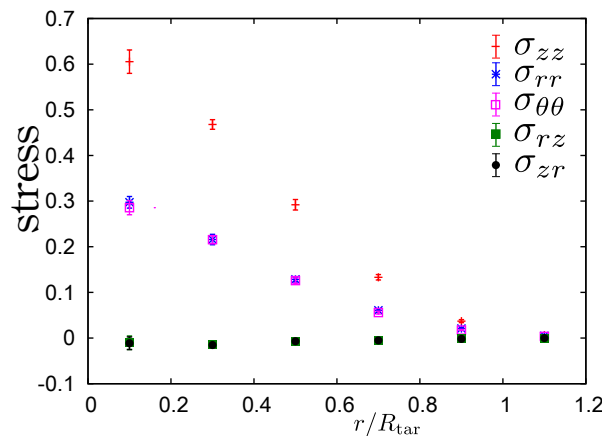}
\caption{(Color online) The profile of the stress tensor $\sigma_{\alpha \beta} (mu_0 ^2 /d^3)$ as the functions of distance from the jet axis $r$ with $R_{\rm tar}$ for frictional grains with $\tilde{\phi_0} = 0.90$. The off-diagonal components of the stress tensor $\sigma_{rz}$ and $\sigma_{zr}$ are much smaller than diagonal components as $|\sigma_{rz}/\sigma_{zz}| \simeq 1.7 \times 10^{-2}$ \ at $r/R_{\rm tar} = 0.1$. }
\label{profile}
\end{figure}

Here we show the profile of the stress tensor for the frictional case (Fig. \ref{profile} for $\tilde{\phi}_0 = 0.90$). 
From Fig. \ref{profile}, it is apparent that off-diagonal components of the stress tensor $\sigma_{rz}$ and $\sigma_{zr}$ are much smaller than diagonal components, where the ratio of the off-diagonal to the diagonal element is estimated as $|\sigma_{rz}/\sigma_{zz}| \simeq 1.7 \times 10^{-2}$\ at $r/R_{\rm tar} = 0.1$. 
This result supports that the solution of Euler equation well reproduce the granular flow after the impact \cite{0}.
We also found that there exists a large normal stress difference, i.e. the difference between diagonal components of $\sigma_{\alpha \beta}$, which is also observed in our two-dimensional case. We obtain the ratio $|\sigma_{rz}/\sigma_{zz}| \simeq 3.0  \times 10^{-2} $\ at $r/R_{\rm tar} = 0.1$ for the frictionless case, where off-diagonal components are much smaller than diagonal ones as in the case of the frictional case.

\subsection{Pressure}
\begin{figure}[h]
\includegraphics[scale = 0.65]{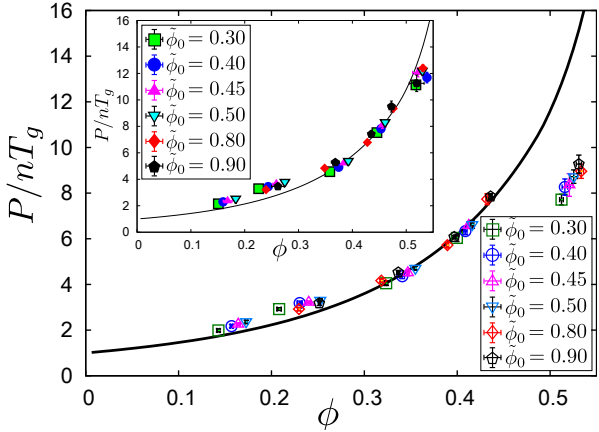}
\caption{(Color online) The comparison between the theoretical pressure in Eq. (\ref{pressure}) and the observed pressure for several $\tilde{\phi_0}$, where the vertical axis is $P$ divided by the density $n$ and the granular temperature $T_g$ for the frictionless case. The inset denotes comparison between those for the frictional case. Black solid lines in each figures denote Eq. (\ref{pressure}) for the frictionless and the frictional case, respectively.}
\label{pressure_fig}
\end{figure}

Let us look at the result of the pressure (Fig. \ref{pressure_fig}). 
It is known that the granular sheared flow such as a chute flow and a plane shear flow can be approximately described by granular hydrodynamics with transport coefficient derived from kinetic theory, i.e. the Enskog equation \cite{5,15,17,18,38,36}. We compare our simulation with the transport coefficients derived by Garz\'{o} and Dufty\cite{36} with the pressure $P\equiv \sum_{\alpha} \sigma_{\alpha \alpha} / 3$, the density $n$, the volume fraction $\phi$, and the granular temperature $T_g({\bf r}) \equiv \sum_{i \alpha} m u^{2} _{i \alpha} /3N$. The pressure is conventionally given by 
\begin{equation}
\frac{P}{nT_g} = 1 + 2\phi(1+e)\chi, \label{pressure} \\
\end{equation}
\begin{equation}
\chi= \left\{
\begin{array}{ll}
\frac{1- \phi /2}{(1-\phi)^3} & \mbox{ ($0 < \phi < \phi_f $) } \\
 \frac{(1- \phi_f /2)(\phi_c - \phi_f)}{(1-\phi_f)^3(\phi_c - \phi)} & \mbox{ ($\phi_f < \phi < \phi_c $),}
\end{array}
\right. 
\end{equation}
where $\phi_f = 0.49$ and $\phi_c = 0.64$\cite{44}. For the frictional case, in general, five equations for rotational degree of freedom are necessary, in addition to those for the translational one.
However, ten equations for frictional grains can be reduced to five equations by introducing effective restitution coefficient $\bar{e}$, if the friction constant $\mu$ is small \cite{15,18,38}. According to this simplification we use the effective restitution coefficient $\bar{e} = 0.616$ for ${e}=0.75$ and $\mu = 0.2$, for frictional case in the following analysis. 

Let us compare the theoretical curve with numerical results for several $\tilde{\phi_0}$ (Fig. \ref{pressure_fig}). The black solid line in Fig. \ref{pressure_fig} and that in the inset denote the theoretical curve for the frictionless case and the frictional case, respectively.
Surprisingly, the expression for the pressure in Eq. (\ref{pressure}) well reproduces the numerical result for $\phi<0.5$ inspite of the existence of the normal stress difference i.e. $\sigma_{zz} > \sigma_{\theta \theta} \simeq \sigma_{rr}$, 
while Eq.(\ref{pressure}) for $0.5<\phi<0.6$ may have significant deviation from the theoretical line. The deviation, which may result from the singularity near the symmetrical axis $r \simeq 0$ of the cylindrical coordinate, emerges only at $r/R_{\rm tar} = 0.1$.

\begin{figure}[h]
\includegraphics[scale = 0.65]{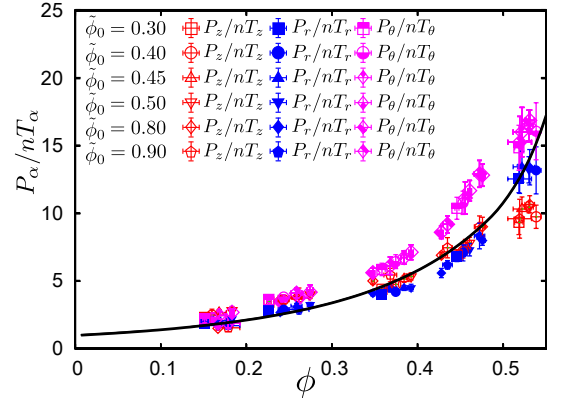}
\caption{(Color online) The diagonal components of the stress tensor for the frictional case with several $\tilde{\phi_0}$ divided by $nT_{\alpha}$, where $T_{\alpha}$ is the temperature for $\alpha$ direction $T_{\alpha} \equiv \sum_i mu_{i \alpha} ^2 /N$. Red empty points, blue filled points, purple half-filled points and the solid black line denote $P_z /nT_z, P_r /nT_r,  P_{\theta} /nT_{\theta}$ and Eq. (\ref{pressure}), respectively.}
\label{anisotropic_press}
\end{figure}

Although there exists large normal stress differences, our numerical results can be reproduced from the empirical relation ($\ref{pressure}$), if we introduce an anisotropic temperature. 
Indeed, equations of state for each coordinate satisfies, $P_{\alpha} = nT_{\alpha} \{ 1 + 2\phi (1+e)\chi \}$ for $\alpha = r, \theta, z$ and $P_{r} = \sigma_{rr}$, $P_{\theta} = \sigma_{\theta \theta}$ and $P_z = \sigma_{zz}$. By summing up $P_{\alpha} = nT_{\alpha}\{1 + 2\phi (1+e)\chi \}$ over $\alpha$, we can reproduce Eq. (\ref{pressure}). 
From Fig. \ref{anisotropic_press}, we verify that $P_z/nT_z$ and $P_r /nT_r$ are on the theoretical curve for isotropic systems, but $P_{\theta}/nT_{\theta}$ has a systematically larger value from the isotropic one. Although our suggestion that the anisotropy of the stress only reflects on the anisotropy of the kinetic temperature is not perfect, the result gives a reasonable physical picture, at least, for $r$ and $z$ directions.

\subsection{Shear viscosity}
Let us evaluate the shear viscosity from the data of the stress tensor. 
The theoretical shear viscosity for frictionless granular fluids, which depends on temperature and volume fraction, is given by
\begin{eqnarray}
\sigma_{rz} &=& -\eta_{\rm kin} D_{rz}  \label{eq_shear}, \\
\eta_{\rm kin} (\phi, T_g) &=& \frac{5}{16d^2}\sqrt{\frac{mT_g}{\pi}} \eta^*(\phi)  \label{shear}
\end{eqnarray}
with strain rate $D_{rz} \equiv (\partial \bar{v}_r/\partial z+\partial \bar{v}_z/\partial r)/2$,
$\eta^* (\phi)\equiv \eta^{k*}[1+4\phi \chi (1+e)/5] + 3\gamma^* /5$, $\eta^{k*}\equiv [1-2(1+e)(1-3e)\phi \chi /5] / (\nu_{\eta} ^* - \zeta^* /2)$, $\gamma^* \equiv 128  \phi ^2 \chi (1+e)(1-c^* /32) /5\pi$, $\nu_{\eta} ^* = \chi [1 - (1-e)^2 /4][1-c^* /64]$, $\zeta ^* \equiv 5 \chi (1-e^2)(1+3c^* /32) / 12$ and $c^* \equiv 32(1-e)(1-2e^2)/[81 - 17e+30e^2(1-e)]$ \cite{36}.

\begin{figure}[h]
\includegraphics[scale = 0.65]{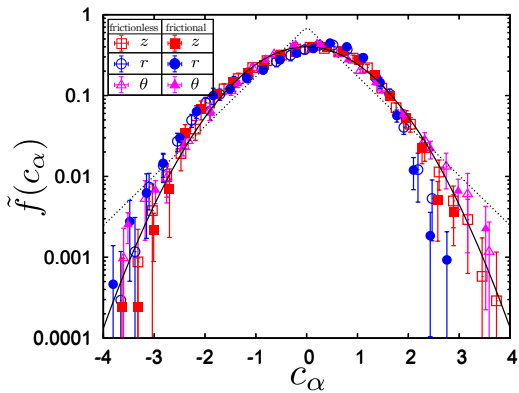}
\caption{(Color online) Scaled VDF $\tilde{f}(c_{\alpha})$ for $\tilde{\phi_0}=0.90$. Empty points and filled points denote the frictionless case and the frictional case, respectively. The black solid line is $\tilde{f}(c_{\alpha}) = \exp(-c_{\alpha} ^2 /2) /\sqrt{2\pi}$ and the dashed line is $\tilde{f}(c_{\alpha}) = \exp(-\sqrt{2}|c_{\alpha}|) / \sqrt{2}$.}
\label{velocity}
\end{figure}

The shear viscosity is usually evaluated by plotting data points on a $\sigma_{rz}$ vs $D_{rz}$ plane. However, as shown in Supplementary Material, if we plot data for each mesh on the plane, the shear viscosity evaluated from a slope on the plane is negative, which is totally unphysical\cite{65}. This negative slope is caused by large density and temperature variations in each mesh. Therefore, the viscosity may be estimated locally by using density and temperature in the corresponding mesh.

For the frictional case, the yield stress $\sigma_Y$, which is the residual stress without deformation may exist in general. 
Thus, the constitutive equation in Eq. (\ref{eq_shear}) is replaced by $\sigma_{rz} = \sigma_Y - \eta D_{rz}$, for this case. 
However, in this paper, we assume $\sigma_Y = 0$. The reason for the absence of the yield stress is based on the following three observations. 

The first reason is the velocity distribution function. 
It is known that the significant effect of Coulombic slip may appear in the non-Gaussianity of velocity distribution functions (VDF), which is characterized by flatness of them. 
For grains in a vibrating container, the VDF are near to the Gaussian for frictionless cases, while the VDF for frictional cases are exponential-like\cite{61}. 
We scale VDF $f(v_{\alpha})$ as $f(v_{\alpha}) = v_{\alpha0} ^{-1} \tilde{f}(c_{\alpha})$ with $\int dc_{\alpha} \tilde{f}(c_{\alpha}) = \int dc_{\alpha} c_{\alpha}^2 \tilde{f}(c_{\alpha}) = 1$, $\int dc_{\alpha} c_{\alpha} \tilde{f}(c_{\alpha}) = 0$ and $v_{\alpha0} \equiv \sqrt{2T_{\alpha}/m}$ for each $\alpha$. Scaled VDF for each velocity components are shown in Fig. \ref{velocity}, where all of the VDFs are near to Gaussian $\tilde{f}(c_{\alpha}) = \exp(-c_{\alpha} ^2 /2) /\sqrt{2\pi}$ even for the frictional case, because friction constant $\mu = 0.2$ is sufficiently small, and are far from exponential-like VDF $\tilde{f}(c_{\alpha}) = \exp(-\sqrt{2}|c_{\alpha}|) / \sqrt{2}$. 
The flatness, which is defined as $\langle x^4 \rangle /\langle x^2 \rangle ^2  = \langle x^4 \rangle $ for $\langle x^2 \rangle = 1$ with $\langle \cdots \rangle  \equiv \int dx \tilde{f}(x) \cdots$, is summarized in TABLE I for $\tilde{\phi}_0 = 0.90$. It should be noted that the flatness with Gaussian VDF is 3.0 and that with exponential VDF is 6.0. 
Although the flatness with $\theta$ component of VDF slightly deviates from $3.0$, it is still far from $6.0$, and thus, the effect of  Coulombic slip with friction constant $\mu = 0.2$ is not significant. 

The second reason is the small Coulombic constant. In this case, renormalization of restitution coefficient is known to be valid\cite{38,18,15}. We stress here that the residual stress for the frictional granular fluid with small Coulombic constant does not exist in a usual setup. We also note that there exists no characteristic feature of Coulombic friction for small $\mu$ such as $\mu = 0.2$ except for the decreases of the jamming density even for the jamming transition\cite{63}.

Moreover, once we assume that shear viscosity corresponds to the value from kinetic theory $\eta(r) = \eta_{\rm kin}(\phi, T_g)$, the extrapolated $\sigma_Y$ are obtained at each mesh, through $\sigma_{rz}(r) = \sigma_Y(r) - \eta_{\rm kin}(\phi (r), T_g(r)) D_{rz}(r)$. As a result, $|\sigma_Y| $ is sufficiently small for $r/R_{\rm tar}>0.40$ (TABLE I\hspace{-.1em}I ). 

 \begin{table}
 \caption{Flatness for the frictionless and the frictional case.}
  \begin{tabular}{c|c|c|c }
                    & z  & r & $\theta$  \\ \hline \hline  
    frictionless case & 2.87 & 2.86 & 3.41 \\  \hline
    frictional case     & 2.70 & 2.98 & 3.71   
           \end{tabular}
\end{table} 

\begin{table}
\caption{Extrapolated yield stress $-\sigma_Y \times 10^{3}$.}
\begin{tabular}{c|c|c|c|c}
    \backslashbox{$\tilde{\phi_0}$}{$r/R_{\rm tar}$}& 0.30 & 0.50 & 0.70 & 0.90\\ \hline \hline  
    0.30 & 6.41 $\pm$ 3.3 & 0.701 $\pm$ 1.5 & 0.434 $\pm$ 0.66 & 0.447 $\pm$ 0.26 \\  \hline
    0.40 & 7.54 $\pm$ 5.8 &-0.159 $\pm$ 3.2 & 0.538 $\pm$ 1.1   & 0.364 $\pm$ 0.52 \\  \hline
    0.45 & 8.79 $\pm$ 2.3 & 0.800 $\pm$ 2.1 & 0.440 $\pm$ 0.94 & 0.783 $\pm$ 0.44 \\  \hline 
    0.50 & 6.79 $\pm$ 2.8 &-0.574 $\pm$ 1.9 & 0.632 $\pm$ 0.94 & 0.608 $\pm$ 0.32 \\  \hline 
    0.80 & 7.00 $\pm$ 4.8 & 2.19    $\pm$ 1.7 &-0.953 $\pm$ 1.3  & 0.392 $\pm$ 0.50 \\  \hline
    0.90 & 5.21 $\pm$ 2.7 &-0.540 $\pm$  2.9 &-0.396 $\pm$ 1.2 & 0.243 $\pm$ 0.83 \\   
\end{tabular}
\end{table} 

\begin{figure}[h]
\includegraphics[scale = 0.63]{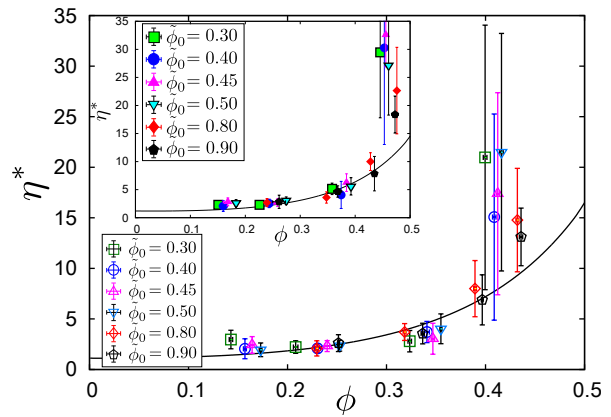}
\caption{(Color online) Non-dimensional shear viscosity $\eta ^*$, which is defined in Eq. (\ref{shear}), for several $\phi_0 /\phi_{\rm fcc}$ in $0.2R_{\rm tar}<r<R_{\rm tar}$. Black solid lines denote theoretical curves.}
\label{viscosity_scaled}
\end{figure}

Thus, $\sigma_Y=0$ is a self consistent assumption if kinetic theory is adopted. 
From these reasons, we assume $\sigma_Y$ = 0.

Now, let us try to compare the theoretical expression in Eq. (\ref{shear}) with the numerical measured shear viscosity. We estimate strain rate as $\partial \bar{v}_r (r,\Delta z/2) / \partial z \simeq ( \bar{v}_r(r,3\Delta z /4) - \bar{v}_r(r,\Delta z /4)) / (\Delta z /2)$ and  $\partial \bar{v}_z (r,z) / \partial r \simeq ( \bar{v}_z(r + \Delta r/2,z) - \bar{v}_z(r - \Delta r/2,z)) / \Delta r$. Since we evaluate the physical quantities near the wall, the mesh $0 < z < \Delta z$ is divided into $0 < z < \Delta z /2$ and $\Delta z /2< z < \Delta z$ to calculate $\partial \bar{v}_r (r,\Delta z/2) / \partial z$ and $0<r<R_{\rm tar}$ is divided into $0<r<\Delta r/2, \Delta r/2<r< 3\Delta r/2, \cdots$. 
The comparison of $\eta^*$, which is the non-dimensional shear viscosity introduced in Eq. (\ref{shear}), for $0.2<r/R_{\rm tar}<1.0$ is shown in Fig. \ref{viscosity_scaled}. Although there is a slight deviation between them for large $\phi$ i.e. small $r$, which may be the effect of the singularity in the center $r=0$, the theoretical curve reproduces other numerical results. We can, thus, conclude that the flow has the finite shear viscosity which has the same order of the predicted value by kinetic theory. The reason why the simulations are approximately described by the Euler equation is that the strain rate itself is small i.e. $0.01 \alt D_{rz} d/ \sqrt{T_g /m} \alt 0.4$, and thus $\sigma_{rz} / \sigma_{zz}$ is small.

\section{Dicussion}
There exists large normal stress differences, while the prediction of the normal stress difference,which appears at Burnett order, is small\cite{64}. We found that the normal stress vertical to the wall is almost twice as large as the other diagonal components of  the stress tensor. A unidirectional flow can be distributed into two directions in the usual time revolution. This mechanism might be easily understood by the time reversal flow in which two directional flow merge into a unidirectional flow. This picture is possible to be considered because the dissipation is not crucially important, at least, for the scattering angle\cite{58}. Thus, the fluctuations in $r$ and $\theta$ components may be half of that in $z$ component. Therefore, we may understand the relation $T_z \simeq 2 T_{\theta} \simeq  2 T_{r}$.

We have assumed the zero yield stress so far, based on the smallness of $\mu$. Here we discuss results for large $\mu$ case. Flatness of VDF for $\mu = 1.0$ and $\tilde{\phi}_0 = 0.90$ is obtained as $\langle c_z ^4\rangle = 3.09$, $\langle c_r ^4\rangle = 3.84$ and $\langle c_{\theta} ^4\rangle = 4.90$, where non-Gaussianity emerges compared with the small $\mu$ case, at least for $\langle c_r ^4\rangle$ and $\langle c_{\theta} ^4\rangle$. We have confirmed that the rotational temperature $R$, which represents the  fluctuation of angular velocity for particles, is within the same order of $T_g$ for $\mu = 1.0$, while $R/T_g \alt 0.2$ holds for $\mu = 0.2$. Since rotational degree of freedom plays a significant role for the large $\mu$ case, theory of dense micro-polar fluid with Coulomb slip, which has not been derived as long as the authors know, would be necessary, to discuss the existence of $\sigma_Y$. The derivation of an explicit description of shear viscosity for granular flow with large $\mu$, as a function of $\phi$, $T_g$ and $R$, is left as a future work.

Hadron physicists calculate a lower bound of $\eta$ via AdS/CFT correspondence, where $\eta$ for QGP is expected to be close to the lower bound\cite{34,32}. Simulation based on a perfect relativistic fluid model well reproduces the behavior of QGP, although QGP may also not be a simple perfect fluid\cite{65}. The similarity between QGP and granular jet reported  in Ref. \cite{1} is superficial, because the shear viscosity is not anomalous, though the shear stress is indeed small.

\section{Conclusion}
We have numerically investigated the granular jet which impacts on a fixed wall. 
We have revealed that the granular flow after the impact has a finite shear viscosity which has the same order as the predicted value from kinetic theory, and thus the similarity between the granular flow and the perfect fluid is superficial, which comes from a small strain rate. This result gives theoretical explanation of the similarity between granular flow and perfect fluid, which is reported in the experiment and the two-dimensional study\cite{1,0}. 
We have assumed $\sigma_Y=0$, judging from VDF, small Coulombic constant and extrapolated $\sigma_Y$. This assumption is strong one for the comparison between kinetic theory and our data. For large $\mu$ case, non-Gaussianity emerges compared with the small $\mu$ case and validity of the renormalization of restitution coefficient may not be valid. 
Results for large Coulombic constant case will be reported elsewhere. 
Although both the pressure and the viscosity are not far from the predictions by kinetic theory, there exists a large normal stress difference in contrast to the case of kinetic theory. 
Our results may shed the light on the internal fluid structure under a strong nonequilibrium situation, i.e. the impact processes of the granular jet.
\begin{acknowledgements}
 We would like to thank W. W. Zhang and T. Hirano for fruitful discussions, and Andrew Hillier for correction during the `English for Scientific Communication' course.
 This work is partially supported by 
and the Grant-in-Aid for the Global COE program ``The Next Generation of Physics, Spun from Universality and Emergenceh from MEXT, Japan.
\end{acknowledgements}


\section*{Supplementary Information: Analysis on a shear stress vs strain rate plane}
In this supplementary material, we try to estimate the shear viscosity data on a shear stress $\sigma_{rz}$ vs strain rate $D_{rz}$ plane. Data points for $\sigma_{rz}$ and $D_{rz}$ in the each mesh for frictional case with $\tilde{\phi}_0 \equiv \phi_0 / \phi_{\rm fcc} = 0.90$, where $\phi_0$ is the volume fraction before the impact and and $\phi_{\rm fcc}$ is that for fcc crystals,  are plotted(Fig. 1). The point for the smallest $D_{rz}$ and for the smallest $\sigma_{rz}$ denote that for $r / R_{\rm tar} = 0.1$ and $r/ R_{\rm tar} = 0.9$ with the target radius $R_{\rm tar}$, respectively. Although from  fitting a line to three other points, we may estimate the shear viscosity, but the estimated value is negative. We have verified that this tendency is insensitive to the choice of a specific $\tilde{\phi}_0$.
\begin{figure}
\includegraphics[scale = 0.50]{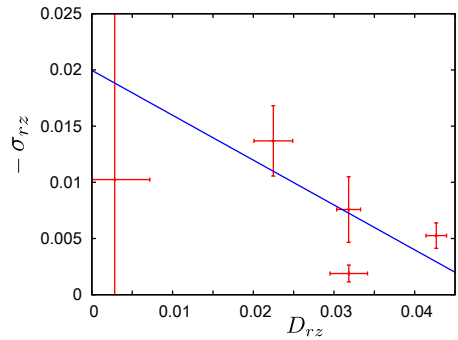}
\caption{(Color online) $\sigma_{rz}$ and $D_{rz}$ for frictional case with $\tilde{\phi}_0 = 0.90$ are plotted. The point for the smallest $D_{rz}$ and for the smallest $\sigma_{rz}$ denote that for $r=\Delta r /2$ and $r = 5\Delta r /2$, respectively. By fitting a line to three other points, negative shear viscosity may be estimated. }
\label{srzdrz}
\end{figure}

\begin{figure}
\includegraphics[scale = 0.55]{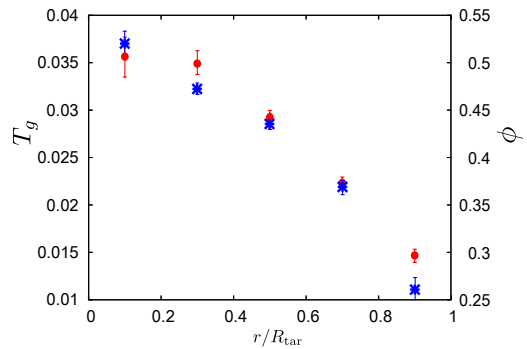}
\caption{(Color online) The profile of the granular temperature and volume fraction. Red points and blue asterisks denote $T_g$ and $\phi$ for the corresponding mesh, respectively.}
\label{temperture}
\end{figure}

The negative viscosity, which is totally unphysical, may be the consequence of the local variation of the volume fraction $\phi$ and the granular temperature $T_g$ between each mesh. 
The profiles of $\phi$ and $T_g$ are shown in Fig. 2. 
In a usual setup, when we estimate the shear viscosity on a $\sigma_{rz}$ vs $D_{rz}$ plane, $\sigma_{rz}$ and $D_{rz}$ are not the local quantities, but the bulk quantities. 
Thus, $\phi$ and $T_g$ are homogeneous and $D_{rz}$ can be controlled\cite{63}. 
However, in our setup with fixing $\phi$ and $T_g$, $D_{rz}$ cannot be controlled. 

Judging from non-uniformity for volume fraction and granular temperature, we do not adopt the viscosity evaluated on the stress-strain rate plane, but adopt the local viscosity as explained in the text.

\end{document}